\documentclass[10pt,conference]{IEEEtran}
\IEEEoverridecommandlockouts
\usepackage{cite}
\usepackage{amsmath,amssymb,amsfonts}
\usepackage{algorithmic}
\usepackage{graphicx}
\usepackage{textcomp}
\usepackage[table,xcdraw]{xcolor}
\usepackage{booktabs}
\usepackage{balance}

\def\BibTeX{{\rm B\kern-.05em{\sc i\kern-.025em b}\kern-.08em
    T\kern-.1667em\lower.7ex\hbox{E}\kern-.125emX}}
\begin{document}

\title{Mining API Interactions to Analyze Software Revisions for the Evolution of Energy Consumption}

\author{\IEEEauthorblockN{Andreas Schuler\IEEEauthorrefmark{1}\IEEEauthorrefmark{2} and Gabriele Kotsis\IEEEauthorrefmark{1}}
\IEEEauthorblockA{\IEEEauthorrefmark{1} Department of Telecooperation
\textit{Johannes Kepler University} Linz, Austria}
\IEEEauthorblockA{ \IEEEauthorrefmark{2}Advanced Information Systems and Technology (AIST)
\textit{University of Applied Sciences Upper Austria} Hagenberg, Austria}
andreas.schuler@fh-hagenberg.at, gabriele.kotsis@jku.at
}

\maketitle

\begin{abstract}
With the widespread use and adoption of mobile platforms like Android a new software quality concern has emerged -- energy consumption. However, developing energy-efficient software and applications requires knowledge and likewise proper tooling to support mobile developers. To this aim, we present an approach to examine the energy evolution of software revisions based on their API interactions. The approach stems from the assumption that the utilization of an API has direct implications on the energy being consumed during runtime. Based on an empirical evaluation, we show initial results that API interactions serve as a flexible, lightweight, and effective way to compare software revisions regarding their energy evolution. Given our initial results we envision that in future using our approach mobile developers will be able to gain insights on the energy implications of changes in source code in the course of the software development life-cycle.
\end{abstract}


\section{Introduction}
Developing sustainable software has become an important issue in mobile software development. Understanding and subsequently reducing the energy consumption of an application has therefore become an essential quality concern for mobile developers. However, developers often lack the expert knowledge as well as the 
tools to improve the energy efficiency of mobile applications \cite{004,950,953,954}.
As a result, the user acceptance for an application that drains
a devices' battery is limited. This was confirmed by several recent studies \cite{527,954,959,004} that revealed that poor battery performance is a mobile users' primary criteria for either purchasing new devices, uninstalling an app or give it poor ratings.

In recent years the research community 
has started to focus on developing methodologies and tools to better understand
the energy implications of software design choices and therefore support 
the development of energy optimized applications \cite{141,518,142,147,011,954,009}.
One key factor in reducing the energy consumption of an application is 
to assess an applications' energy consumption and attribute the recorded
energy profiles back to program structures at a fine granularity, e.g. method-level.
Throughout the years several approaches have been described to tackle
this challenge, either by using power modeling \cite{142,214}, measurement devices \cite{014} 
or by interfacing with a devices' hardware counters, e.g. using Intel's RAPL interface \cite{957,009}.  

However, obtaining an energy profile is a challenging task, as it takes time to record the relevant data and map it back 
to individual program structures. Therefore, researches have spent effort towards 
finding proxies which help to explain the energy consumption without the need for laborious 
profiling. Researches have therefore successfully shown a connection between 
the frequency of system calls and energy consumption \cite{217,533,142,147}, have proposed 
an approach to recommend energy-efficient Java collections \cite{913} or did an in-depth investigation on the energy profiles of Java collection types \cite{912}. 
Furthermore, the energy implications of Java IO classes 
\cite{009}, logging frameworks \cite{31} and the effects of bundling API calls to achieve better energy performance \cite{954} have been investigated.  

Our work contributes to this field of research, however, in contrast
to the existing work, in this paper we present an approach to examine the evolution of the energy consumption of a library through mining its API interactions. This way, we are able to give feedback to the developer in the course of the software development life-cycle, whether a change in code led to an in- or decrease of the energy consumption. 
As a result, our work ultimately supports developers to gain insights about 
the energy implications and take corresponding actions already during development.
This paper significantly differs from our previous work \cite{956}, as 
the objective is to assess the applicability of a previously defined model
as a proxy for the evolution of energy consumption in software revisions in the context of mobile software development. Consequently, this paper makes the following contributions:

\begin{itemize}
  \item
    An approach to obtain and assess the difference in energy consumption between software revisions by analysing its runtime API interaction behaviour.
  \item
    Empirical evidence on the effectiveness of the presented approach to
    claim the evolution of a software artifact's energy consumption. 
  \item
    A replication package consisting of the recorded data based on
    the analysis of software revisions of a publicly available library over the course of 10 years \cite{anon}.
  \end{itemize}

\section{Background}\label{sec:background}

\subsection{Software Energy Consumption Profiling}
Software energy profiling is defined as to estimate the consumed energy based on
energy/power models \cite{003}. Based on the kind of input variables
that are used to define such a model, power modeling can be categorized
in 3 groups: \emph{utilization-based}, \emph{event-based} and
\emph{code-analysis-based} models \cite{003}. 
A utilization-based model is defined based on the resource utilization of
specific hardware components. An event-based model is the result of
correlating system events like system-calls \cite{217}, 
API interactions \cite{956} or sequences of API calls \cite{179} with energy readings. 
Finally, a code-analysis-based model allows to estimate the energy
consumption based solely on source code inspection \cite{003}.

In recent years several approaches have been established
to estimate the energy consumption of an application, for each 
of the categories stated above. This includes mining
for energy inefficient API patterns \cite{179}, recommending 
energy efficient Java collections \cite{913,912} as well as 
comparing the energy characteristics of Java IO approaches \cite{009}. 
Besides that, there is a growing body of work which examines
the impact of software design on the energy consumption \cite{011}. 
Additionally, Intel's RAPL interface has gathered a lot of research interest and is widely used to derive software energy profiles \cite{957,009}.

%\begin{figure}[h]
%    \centering
%    \includegraphics[width=0.40\textwidth]{images/%experiment_def.pdf}
%    \caption{High level architecture of the %\textsc{Mana} platform and structure of our %experiment test-bed. The \emph{Experiment %Execution Coordinator (EEC)} component 
%    is connected to the Power Monitor and the %Android device, adapted from \cite{013}.}
%    \label{fig:expdef}
%\end{figure}

For the experiments carried out in this paper and to obtain energy profiles, we rely 
on measurements obtained using the Monsoon High Voltage Power Monitor (HVPM) \cite{005}, a device that is commonly used for energy consumption analysis in mobile ecosystems \cite{956,187,120,214}.
In particular, we use the Monsoon power monitor as part of our test-bed introduced in Schuler and Kotsis \cite{014}. The test-bed serves as an environment to execute tests, record the energy profiles at a sampling rate of 20 Khz and map the resulting profiles back to program structures.

\subsection{A Model based on API Utilization}
The API utilization metric or in short $U_{api}$ is a notion of
how a particular method in a library or application 
interacts with a provided API. This metric has its foundation in our previous work \cite{956}. However, for enhanced replicability we give a description of the model as follows:
The metric stems from the assumption that the way a method interacts
with a provided API has direct influence on its 
energy characteristics \cite{956}.

The metric is computed
by obtaining the dynamic call graph from a library or application under test. 
Using said call graph, all nodes are traversed and the frequency of API interactions for each method being called is recorded. As API interaction we define any method call from a particular library or application under test into the Java-based Android Platform API. Henceforth, based on a dynamic call graph $CG(V,E)$, with vertices $V$ for each method being executed and 
edges $E$ denoting an API call from one method to another,
$ad(v)$ is defined as all methods that are being called 
from method $v \in V$. Thus, $ad(v)$ returns all vertices 
in $CG$ which have an incoming edge $e \in E$ from $v$. 
The formal definition of $ad(v)$ reads as follows:

\begin{equation}
  ad(v) := \{ u  | v, u \in V \land (v, u) \in E \}
\end{equation}

Using $ad(v)$ API utilization $U_{api}$ of given method $v$ is defined as:

\begin{equation}
  U_{api}(v)= 1 + \sum\limits_{i=1}^{\mid ad(v) \mid} U_{api}( ad(v)_i )
\end{equation}
With $\mid ad(v) \mid$ being the cardinality, hence the number of API interactions issued by $v$. Methods with no API interactions have a $U_{api}$ value of 0. 
In addition to our previously proposed definition \cite{956} of $U_{api}$, we suggest an adaption to the metric by dividing the computed $U_{api}$ value by the number of total API interactions $N$ and thus get the relative
$rU_{api}$, with $N$ denoting the sum of all API interactions for a particular library under test:

\begin{equation}
\label{eq:relative}
    rU_{api}(v) = \frac{U_{api}(v)}{N + 1} 
\end{equation}

\begin{figure*}[!t]
  \centering
  \includegraphics[width=\textwidth]{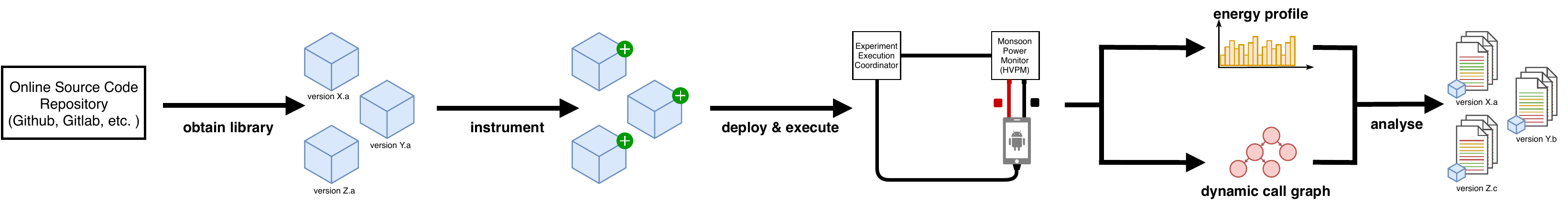}
  \caption{Approach to assess the API profile and the energy characteristics partly composed of our energy test-bed introduced in \emph{[omitted for review]}.}
  \label{fig:approach}
\end{figure*}

\section{Approach}\label{sec:method}
Figure \ref{fig:approach} presents our approach which 
aims to support developers in the course of the software development life-cycle 
in order to determine if a change in code results in an alteration 
of the energy consumption of the artifact being analysed. Furthermore,
the proposed approach builds the foundation to assess the change 
in energy consumption between different software revisions and thus
enables developers to take timely counter measures in order
to decrease the overall energy.

Our approach is structured in 
a series of consecutive steps. 
The initial step, as depicted in Figure \ref{fig:approach} consists of  
selecting and obtaining a software artifact from a version control system.
Next, the approach extracts all revisions of the artifact from the respective repository.
Each revision is further instrumented to collect runtime traces 
for when the artifact is being executed. This is achieved by locating all available unit tests in the library and inserting probes to collect the call trace during execution using the Android Debug class.

Once an artifact is instrumented, it is wrapped in an Android application package, and deployed to the test-bed described in section \ref{sec:background}. 
The unit tests are executed on the Android device. For each test 
we record its energy profile using the Monsoon HVPM as well as its call trace. To account for possible outliers in the recorded energy profiles, each test execution is sampled and energy data is averaged.

Using the collected call traces, a dynamic call graph is constructed
per test method and the $rU_{api}$ metric is computed for each obtained graph. By applying the $rU_{api}$ computation based on the packages the methods being called reside in, it is possible to derive
a distribution of the API interactions per method. Additionally, we record package, class, method name and thread-id 
as well as start and duration of all methods being called along the call trace.
Finally, the recorded energy profile is attributed to individual methods, by 
determining start and end in the energy profile relative to the recorded
method attributes obtained via the collected call traces.

The final result, as depicted in Figure \ref{fig:approach},
constitutes the executed methods each with attributed energy consumption
and computed $rU_{api}$ profile.
The collected data serves as a foundation for further analysis on the evolution 
of a software artifact with respect to its Android Platform API interactions and the resulting implications on its energy characteristics.

\section{Empirical Evaluation}\label{sec:evaluation}
By applying the presented approach to a software
artifact, we aim to answer, if it is feasible
to detect changes in energy consumption between
software revisions by solely examining the differences
in the API interactions. To achieve this, 
we carry out an empirical study based on 
a selection of software revisions taken from an open source
library for JSON document processing.

\subsection{Research Questions}
Using the presented approach we seek to answer the following research questions ($RQ$):
\begin{itemize}
  \item
  \textbf{\(RQ_1\):} \emph{How does the energy consumption of Google Gson library evolve between software revisions?} Whether it is new features that are being added or simply by fixing a reported bug, a software's code base is subject to regular change. In order to determine how such a change affects its energy footprint, we seek to investigate, how the energy consumption differs amongst selected software revisions of Google Gson library.
  \item
  \textbf{\(RQ_2\):} \emph{How can a change in energy consumption be 
  inferred between different software revisions by analysing its API interactions?}
  By applying the defined metric in section \ref{sec:background} to different 
  software revisions of a library examined, we want to verify whether 
  it is possible to infer how a change in the API profile affects the energy consumption
  by looking at the evolution of the $rU_{api}$ profile only. 
\end{itemize}

\subsection{Studied Sample}
We based our study on the Google Gson library, a library which is 
commonly used for JSON document processing in mobile ecosystems.
From its GitHub repository we obtained released revisions 
over the course of 10 years, starting from 2009 till 2019. In total 14 revisions
were collected which serve as an input for our approach.
We focused on major and minor revisions solely and discarded revisions only
consisting of bug-fixes and patches. For an overview on the selected revisions,
refer to Figure \ref{fig:energy}. Details about the selected revisions can be found in the description provided as part of the replication package \cite{anon}.
As we base our analysis on the available unit tests, we selected a sub
set of the available tests, to ensure each test being analysed is present
in every revision. This is due to the fact that the number of available test cases in the examined library rapidly grows between revisions, which would lead to an uneven number of sample sizes, and therefore reduce statistical power of the applied analysis of variance.
Hence, we selected the 100 most energy demanding tests and evaluated their presence in all of the 14 revisions which ultimately results in a curated list of 41 unit tests. 

\subsection{Methodology}

To answer \(RQ_1\) we record the energy consumption
for each library revision being examined using the
approach presented in section \ref{sec:method}. To determine how
the energy consumption of a particular library evolves
between the different revisions, we record its
average wattage in milliwatts, energy consumption in joules and duration in milliseconds over the set of selected tests. In order to account for possible outliers we sample test executions 10 times. Finally, we
apply statistical analysis and compare the energy characteristics of examined revisions for significant changes.

To answer \(RQ_2\) we additionally compute the $rU_{API}$ for each
of the selected unit test methods. By computing the
profiles based on the selected unit test methods, each method
is attributed with a distribution of the API utilization, thus
providing an insight on how the examined revisions and the selected
tests interact with the Android Platform API, respectively. For
a detailed overview on the covered APIs, refer to our replication 
package \cite{anon}. We further apply statistical analysis to the
obtained data to evaluate if a significant change in the API profile
of a library revision is also reflected in a significant change in 
energy consumption. First, we group the recorded and attributed unit tests 
by their library version. Second, we apply one way ANOVA individually for 
energy consumption, average power and computed $rU_{API}$ values between all selected revisions. Finally, we apply a post-hoc Tukey test to determine, if
changes in energy consumption, average power and $rU_{API}$ values amongst revisions are significant. Given the results obtained this way, we further compute the accuracy of using $rU_{API}$ as a proxy to determine a significant change in energy and power consumption, respectively.

\section{Results}\label{sec:results}
\begin{figure}[!h]
  \centering
  \includegraphics[width=0.48\textwidth]{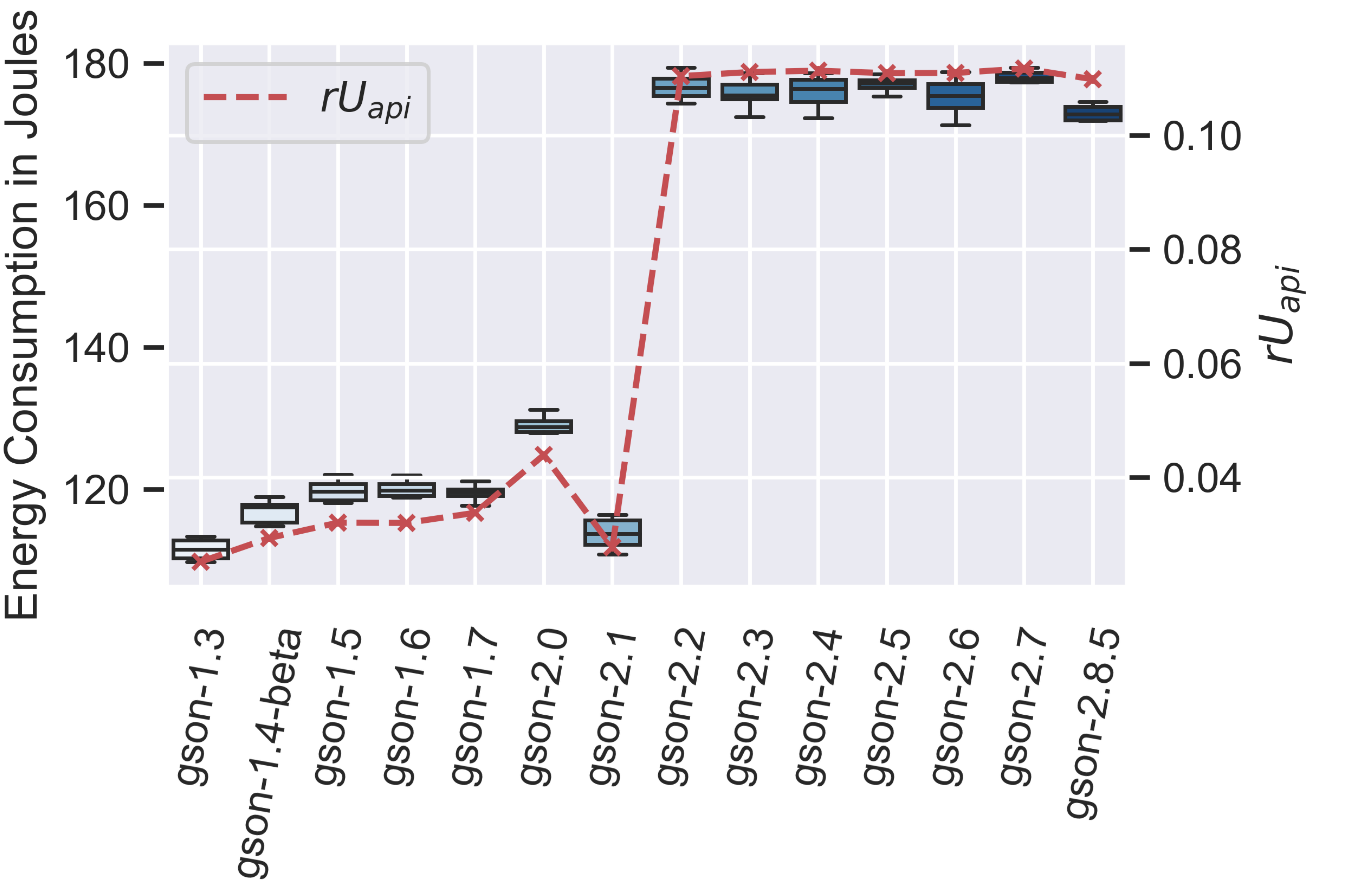}
  \caption{Comparison of $rU_{API}$ profiles and energy consumption for the studied samples.} 
  \label{fig:energy}
\end{figure}

In total, we recorded $41$ test executions per each of the $14$ revision. Each test was further sampled $10$ times which results in a total of 5740 recorded test executions. For answering research question \(RQ_1\), 
we analysed the selected software revisions for their average power usage and energy consumption. For each library we instrumented the selected tests, deployed it to the 
test-bed and collected their average power, their energy consumption as well as their runtime. We further computed the total average power and average energy consumption per examined revision. Referring to Figure \ref{fig:energy}, we can observe that starting from revision 1.3 the trend in the energy consumption is slightly increasing between versions. However, between revision 2.1 and 2.2. the energy consumption recordings show a substantial rise. To investigate the cause for this substantial rise, we compared the changes of revisions 2.1 and 2.2 using git diff command. Apparently, in version 2.2 a new class \textit{com.google.gson.internal.StringMap} was added. The class was created as a replacement for the \textit{java.util.LinkedHashMap} implementation which is part of the Android Platform API. Our assumption is that it is likely that \textit{StringMap} is responsible for the observed behaviour. However, to claim if our assumption is correct, we need to carry out further experiments, which we leave to future work.

\begin{figure}[!t]
  \centering
  \includegraphics[width=0.48\textwidth]{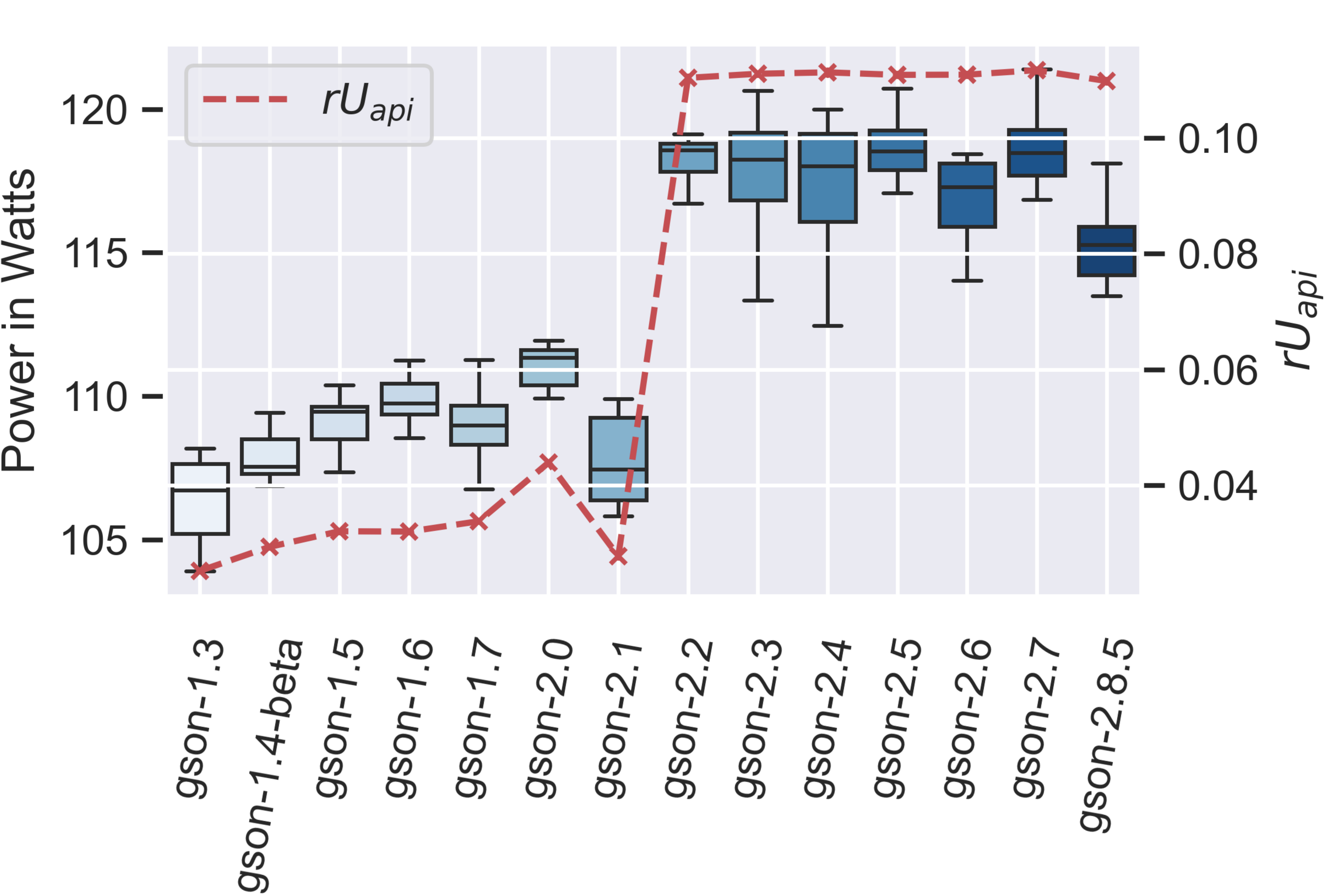}
  \caption{Comparison of $rU_{API}$ profiles and average power consumption for the studied samples.}
  \label{fig:power}
\end{figure}

To answer \(RQ_2\) we seek to investigate if by solely examining the API interactions of a particular version, we are able to infer the 
evolution of the energy characteristics of a revision. To achieve this, besides the recorded energy attributes and the computed the $rU_{api}$ for each test method selected, we computed the sum of $rU_{api}$ per revision. Let alone from visual inspection we can observe that the API utilization follows the trend of both the energy consumption (cf. Figure \ref{fig:energy}) and the average power (cf. Figure \ref{fig:power}). We applied analysis of variance (ANOVA) which yields a significant variation among the revisions, regarding their energy consumption ($F=2527.74,p<.001$), their average power ($F=93.56,p<.001$) as well as their $rU_{api}$ values ($F=24997.37,p<.001$). 

We further applied a post-hoc Tukey test to examine the differences between revisions for statistical significance. We compared the results from the post-hoc test for $rU_{api}$ with the scores obtained for the average power and energy consumption. Henceforth, we computed the accuracy of the presented approach as the ratio between cases that were determined correctly using the presented model for API interactions in relation to all available cases. A correctly determined case is considered a case where the API profile is able to correctly state that the energy characteristics stay the same or significantly change between compared revisions. Consequently, our model for API interactions, when compared with the energy consumption, shows an accuracy of $85\%$. When comparing the model with the recorded average power it still reaches an accuracy of $74.7\%$ percent. Due to the fact that the examined data is not symmetrically distributed, e.g. there are more cases where there is a change in energy consumption or recorded power compared to cases where there is no change, we additionally report the F1-score for energy consumption ($F1=0.91$) and power ($F1=0.82$).

Given the results, we are confident that the applied model for API interactions can serve as a simple approach to assess the evolution of energy consumption amongst software revisions for libraries and applications utilizing CPU, memory and I/O bound APIs. However, we cannot draw a definitive conclusion as we evaluated the presented approach solely using one library. To account for this threat to its validity, for future work we want to extend our initial research approach and apply it to more libraries, with different Android Platform API interaction characteristics.
\section{Conclusion}\label{sec:conclusion}
In this paper we introduced an approach
to mine the API interactions of software revisions and use
respective data as a proxy for the evolution of energy consumption.
Our results show that our approach is both effective and lightweight
in nature, as it does not involve any alterations on the target device.
We believe that the presented approach, when applied as part of a development
environment, can support developers in order to make proper design decisions 
with energy consumption in mind. For future work, we want to examine which change could be responsible for an in or decrease, 
especially with cases where there is a substantial rise between revisions, as depicted in Figure \ref{fig:energy} between Gson 
version 2.1 and 2.2. Besides that, we are currently focusing on including weights in the $U_{api}$ metric computation to better 
account for the attribution of different APIs to the overall metric.

\bibliographystyle{IEEEtran}
\balance
\bibliography{IEEEabrv,literature}

\end{document}